# Inertial Oscillations of Pinned Dislocations

M. Bhattacharya[1], A. Dutta[2], P. Mukherjee[1], N. Gayathri[1] and P. Barat[1]

[1]*Variable Energy Cyclotron Centre, 1/AF, Bidhannagar, Kolkata 700 064, India*

[2]*Department of Metallurgical and Materials Engineering, Jadavpur University, Kolkata 700032, India*

**Dislocation pinning plays a vital role in the plastic behaviour of a crystalline solid. Here we report the first observation of the damped oscillations of a mobile dislocation after it gets pinned at an obstacle in the presence of a constant static shear load. These oscillations are found to be inertial, instead of forced as obtained in the studies of internal friction of solid. The rate of damping enables us to determine the effective mass of the dislocation. Nevertheless, the observed relation between the oscillation frequency and the link length is found to be anomalous, when compared with the theoretical results in the framework of Koehler's vibrating string model. We assign this anomaly to the improper boundary conditions employed in the treatment. Finally, we propose that the inertial oscillations may offer a plausible explanation of the electromagnetic emissions during material deformation and seismic activities.**

Dislocations are line defects mediating the plastic deformation, which has immensely driven the development of human civilisation since the onset of the Chalcolithic period. A moving dislocation can get pinned upon its interaction with obstacles like point defects, voids, precipitates and other dislocations[1-5]. Particularly during strain hardening, dislocation pinning becomes a frequent event resulting from a reduction in the effective mean free path of the mobile dislocations[6,7] and its kinetic energy gets liberated when the dislocation comes to rest. On one hand, dislocation pinning dictates the mechanism of plastic deformation in both the ductile and brittle modes, while on the other, it is a compulsory ingredient of the phenomenon of internal friction of solids[8,9]. In case of plastic deformation, it is generally believed that a dislocation spends most of the time at the pinning site until the local resolved shear stress exceeds the critical depinning stress for that particular pinning obstacle. As in the general theory of kinetics, the slowest step is regarded as the rate controlling step, the thermodynamic model of dislocation mediated plasticity[10] considers the pinning-depinning process to be a major factor governing the rate of plastic deformation. In this context, researchers have primarily focussed on the process of depinning and its relation with the nature of the obstacles[3,11-13]. Similarly, in the studies of internal friction, dislocation pinning has a twofold role. The strong obstacles serve the purpose of clamping the ends of the dislocation segment whereas, the weaker obstacles provide an extra drag force on the vibrating dislocation segment[14,15]. It is important to note that in neither of the plastic deformation and the internal friction studies, the exact mechanism of pinning of a mobile dislocation has drawn the deserving attention of the dislocation physicists. Maintaining a sceptical outlook on this lack of interest, we intend to study this phenomenon of dislocation pinning, down to the atomistic level. Such an understanding at the ultrafine scales of length and time is not possible to be gained from the laboratory experiments and the *in silico* methods are apparently the most fruitful in this case. As presented below, our simulation results disclose novel features pertaining to dislocation science, which hitherto remained obscured.

We use an edge dislocation in b.c.c. crystal as a model system in our simulations. Figures 1a-f show the molecular dynamics (MD) simulation snapshots of the pinning of the moving dislocation in molybdenum by a nanovoid that acts as an obstacle (see the Methods section for details of simulation procedure). A dislocation with an effective mass $M$ attains the kinetic energy $\frac{1}{2}M\upsilon_0^2$ ($\upsilon_0 = \frac{\tau b}{B}$ is the terminal velocity with Burgers vector $b$, shear load $\tau$ and drag coefficient $B$) before its motion is suddenly hindered by the obstacle. It is interesting to note that instead of losing all of its kinetic energy instantly, the pinned dislocation segment is found to exhibit oscillatory dynamics under the constant applied shear load of 50 MPa, which is less than the critical depinning stress for the present simulation parameters. For the sake of clarity, the $y$ coordinate of the centre of mass of the dislocation core atoms is displayed as a function of elapsed time in Fig. 1g, where one can easily recognise that the dislocation segment comes to rest by means of damped oscillations. Oscillations of pinned dislocations have been a common phenomenon encountered by dislocation physicists during the studies of internal friction and ultrasonic attenuation experiments[16,17]. These are essentially forced oscillations of dislocations induced by the external cyclic loads. Surprisingly, what we report here is the first observation of the spontaneous oscillations of a pinned dislocation segment under the continuously applied static load. At this point it is worthwhile to study how a static load can induce oscillatory behaviour of dislocations. Moreover, what are the factors responsible for deciding the frequency of oscillations and the rate of damping? We further investigate this strange observation through analytical modelling.

Koehler[18] introduced the string model for pinned vibrating dislocations (further developed by Granato and Lücke[19]) to investigate their role in the internal friction of solids. Motivated by the wide acceptance and success of this model, we apply it for interpreting the observed spontaneous oscillations theoretically. According to this model, the dynamics of a pinned dislocation segment with its line along the $x$-direction can be described as,

$$M\frac{\partial^2 y}{\partial t^2} + B\frac{\partial y}{\partial t} - T\frac{\partial^2 y}{\partial x^2} = \tau b \quad . \tag{2}$$

Here $y(x,t)$ is the dislocation displacement along the $y$-direction at a time $t$ and $T$ is the line tension. In the framework of the MD simulation scheme, we assume that the moving dislocation line is incident on the obstacles with velocity $v_0$ and $l$ is the distance between two adjacent obstacles (the link length). Consequently the boundary conditions are

$$y(0,t) = 0, \quad y(l,t) = 0 \tag{3}$$

and the initial conditions are

$$y(x,0) = 0, \quad \frac{dy(x,0)}{dt} = v_0[u(x) - u(x-l)], \tag{4}$$

where $u(x)$ stands for the unit step function. Under the given boundary and initial conditions, the auxiliary equation produces the solution for the centre of mass position of the string as (see the Supplementary Note 1),

$$y_{cm}(t) = \sum_{m=0}^{\infty} \frac{8v_0 l}{(2m+1)^2 \pi^2 \sqrt{(2m+1)^2 \pi^2 c^2 - \eta^2 l^2}} e^{-\eta t} \sin\left(\sqrt{\frac{(2m+1)^2 \pi^2 c^2}{l^2} - \eta^2}\right) t, \tag{5}$$

where $c = \sqrt{\frac{T}{M}}$ and $\eta = \frac{B}{2M}$. Correspondingly, the first harmonic is the most prominent frequency component and is given by $y_{cm}(t) = \frac{8v_0 l}{\pi^2 \sqrt{\pi^2 c^2 - \eta^2 l^2}} e^{-\eta t} \sin\left(\sqrt{\frac{\pi^2 c^2}{l^2} - \eta^2}\right) t$. Equation (5) clearly shows that the dislocation segment executes damped oscillations driven by the initial momentum of the moving dislocation even under non-cyclic load and hence the oscillations are inertial. The term $e^{-\eta t}$ reflects the dissipative work done by the frictional forces of the medium on the string; in other words, the kinetic energy of the dislocation gets dissipated in the form of heat. We find that a larger ratio of the drag coefficient to the effective mass causes faster damping with smaller oscillation amplitude and the dynamics will be practically overdamped at sufficiently high temperatures. The oscillation profile shows a perfect exponential damping, which proves to be a useful tool to determine the effective mass of the dislocation. In our case, the effective mass is estimated to be $3.6 \times 10^{-16}$ kg/m.

Having found that equation (5) establishes the damped inertial oscillations of dislocation during deformation, it is crucial to investigate the fitness of the employed mathematical formalism in dealing with the actual underlying dynamics. Right hand side of equation (2) is a constant value and thus the complementary function merely decides the phase of the oscillations without affecting the frequency. Equation (5) indicates that the fundamental oscillation frequency scales with the link length as $f_0 = \frac{1}{2\pi}\sqrt{\frac{\pi^2 c^2}{l^2} - \eta^2}$. To verify whether the Koehler's framework can accurately describe the observed oscillation, the link length $l$ is varied between 3.9 - 23.1 nm by altering the simulation cell dimension along the $x$-direction. We perform the Fast Fourier Transform (FFT) of the oscillation data of the MD simulation to evaluate the fundamental oscillation frequencies $f_0$, as demonstrated in Fig. 2a. Figure 2b represents the link length dependency of the extracted frequencies in terms of the $f_0^2$ vs. $l^{-2}$ plot. Although equation (5) predicts a linear relation between $f_0^2$ and $l^{-2}$, we can notice a considerable extent of deviation from this behaviour in the MD results. Curious about the anomalies in the frequency plot, we explore if there is a room for appropriate modification in the existing formalism.

The $f_0^2 \propto l^{-2}$ relation is the manifestation of the fact that the end points of the vibrating string are ideal i.e. perfectly clamped. Thus, the boundaries are completely reflective (with no transmittance) thereby creating standing waves with natural frequency ($f_0 = \frac{\pi c}{l}$) following a reciprocal relation with the link length $l$. A deviation from this behaviour is the signature of partial transmittance of the boundaries, which signifies some extent of flexibility of the clamped ends. In presence of such flexible clamping, the oscillation of a clamped string segment remains no more independent; rather, it gets coupled with the vibrations from the other segments. In order to depict this concept analytically, we propose to realize the imperfection of the boundaries by modelling the endpoints as clamped to hookean springs with stiffness constant $k$ as shown in Fig. 3a. We consider an array of springs clamped with the string at positions periodically separated by the gap $l$. Assuming identical initial conditions for each of the string segment, the dynamics of each segment would also be identical. In this case, the Koehler's equation of motion is applicable with periodic boundary conditions as

$$y(0,t) = y(l,t), \quad T\frac{\partial y}{\partial x}\bigg|_{x=0} - T\frac{\partial y}{\partial x}\bigg|_{x=l} = ky(0,t), \quad (6)$$

along with the initial conditions

$$y(x,0) = 0, \quad \frac{dy(x,0)}{dt} = v_0. \quad (7)$$

The boundary conditions (6) lead to different $f_0^2$ values, significantly deviated from the $\sim l^2$ relation. The ideal dependence (between $f$ and $l$) for the perfectly rigid boundary originates from the relation $pl = n\pi$ (from equation (3)), where $p$ is the wavevector and $n$ is a positive integer. However, in case of flexible boundaries, equations (6) yield (see the Supplementary Note 2)

$$\Gamma(l,p,T) = 2Tp \ \tan\frac{pl}{2} = k, \quad (8)$$

which is nonlinear and an exact solution cannot be derived. We therefore plot its numerical solution in Fig. 3b, where the allowed solutions are shown in $\Gamma$–$p$-$l$ plot. To elucidate the physical implications of flexible ends, we study the dynamics of the string segment with different spring constants by numerical method (see Methods section for details). This method successfully produces the damped oscillation profiles to give the corresponding frequencies as shown in Fig. 4a. We produce $f_0^2$ vs. $l^2$ plot in Fig. 4b for the spring constants $k=$ 5, 10 and 20 to facilitate a direct comparison with Fig. 2b. The result corresponding to the perfectly rigid clamping is also presented as a reference in the inset. As expected from equation (5), the reference plot is a perfect straight line, whereas all the other curves in Fig. 4b deviate from the linear behaviour. A spring having smaller stiffness is found to cause a larger deviation from the fixed boundary behaviour. In particular, for large extent of coupling, i.e small value of spring constant, equation (8) can be approximated as $2Tp \tan\frac{pl}{2} \approx Tp^2l = k$, which gives the smallest and dominant frequency component $f_0 = \frac{1}{2\pi}\sqrt{p^2c^2 - \eta^2} = \frac{1}{2\pi}\sqrt{\frac{k}{lm} - \eta^2}$. We find that the modifications in the boundary conditions have become extremely significant as they reduce the $f_0^2 \propto l^{-2}$ relation to a drastically different $f_0^2 \propto l^{-1}$ expression. We confirm this by rescaling the abscissa of the Fig. 4b from $l^{-2}$ to $l^{-1}$ to obtain a linear behaviour for $f_0^2$ (refer Fig. 4c). Surprisingly, this linear behaviour is reproduced even for the MD results as presented in Fig. 4d. This provides a firm ground for expecting a sufficiently large coupling of oscillations, quantifiable in terms of the spring constant. A linear fit to the MD output can estimate the effective spring constant using the exact expression of $f_0$ given above and we obtain $k=15.4$ N/m for our simulation system. This physically signifies a spring of very small stiffness, which justifies the observed linear relation between $f_0^2$ with the link length. One can now figure out that the trend in Fig. 4d, as obtained from the atomistic simulations reveals that the dynamics should be more appropriately described in the architecture of flexible boundary model, instead of the conventional one.

There is still an important issue to be resolved. We know that Koehler's equation is justifiable for a pinned dislocation because of its direct correspondence with a clamped string under tension. The linear density of the string is analogous to the effective mass of the dislocation, the coefficient of viscosity corresponds to dislocation drag[20] and the string's tension to the elastic energy per unit length increment of the dislocation line. As the MD simulation output highlights the appropriateness of the flexible boundary conditions, physical interpretation of the *spring* introduced in our formalism becomes essential. Introduction of springs in the string model makes the boundaries movable. However, it is worthwhile to emphasize that one should not expect the motion of the obstacles in the MD simulations. Exactly what happens here is that each oscillating pinned dislocation segment (primary and all periodic segments) acts as an emitter as well as receiver of the elastic waves. Such an oscillator interacts with the elastic waves emitted by the other oscillators present in the system. Owing to the finite velocity of these waves, the oscillator may be out of phase with a wave incident on it. This phase mismatch manifests in a work done by the specified pinned segment under discussion, which is analogous to the work done against the restoring forces of the springs depicted in the model. The coupling of oscillations invoked through the springs is physically understandable as the coupling realized through the emitted elastic waves. The observations discussed so far verify that individually oscillating dislocation segments interact with one another instead of oscillating independently. This conclusion is significant not only in the case of inertial oscillations, but for the forced oscillations considered in the internal friction related studies as well. It can be noted that the present analysis assumes equally spaced springs to make it compatible with the periodic boundary conditions of MD simulations. Nevertheless, the concept of interacting oscillations of the pinned dislocation segments would be equally valid for a distribution of link length and orientation in realistic circumstances.

So far, we have kept assuming that the entire kinetic energy recovered after a moving dislocation gets pinned, is released to heat up the crystal. The oscillating dislocation segments interact with the drag forces, primarily the phonon drag, to bring about this effect. But is it the only possible mechanism for dissipation of the dislocation's kinetic energy? As we know by now that dislocations show oscillatory dynamics after pinning, an additional mode of kinetic energy dissipation can be put forward. It emerges when the electrical charge density of the crystal is taken into account. The translational invariance of the crystal structure along the direction of a straight dislocation line ($x$-direction) causes the electrical charge density to vary in the same direction with a period of interplanar ($yz$-plane) distance. However, this periodicity is not maintained if the dislocation line bows out between two pinning sites under stress. In case of inertial oscillation as observed in MD simulation, the charge density would also oscillate periodically with time. Such periodic time variance of electrical charge density creates electromagnetic field that propagates in space carrying a power density quantified in terms of its Poynting vector. As this modality of energetics is dissipative in nature, the electromagnetic emission is expected to impart an extra drag force to the dislocation. In contrast to the other known dislocation drag mechanisms, this electromagnetic drag is peculiar in the sense that it would come into picture only for a vibrating dislocation line. It offers a plausible explanation of the observations of emission of electromagnetic waves in the laboratory experiments of material deformation[21-25] and during the earthquakes[26-28]. The reason behind these emissions has remained obscured and debatable till date. The measured electromagnetic signals are found to be more prominent in alloys and impure metals as compared to metal of high purity[21], which is supportive to our view as the inertial oscillations are less probable in pure materials having a smaller concentration of pinning sites.

The present simulations reveal the occurrence of inertial oscillations of pinned dislocations in crystalline solids. The MD simulations, theoretical analysis and the numerical implementation are in excellent agreement to conclusively prove that oscillations of pinned dislocation segments are indeed coupled and further work on this aspect may need to treat this coupling in terms of the direct elastic field theory. The flexible boundary scheme introduced here calls for a revisit of the theory of internal friction in solids. Finally, dedicated studies must be carried out to verify the relation between the inertial oscillations and the electromagnetic emission during material deformation.

**METHODS**

**Molecular Dynamics Simulation.** The MD simulations in this work are performed using the MD++ molecular dynamics package[29]. The simulation supercell holds the b.c.c. crystal of Molybdenum realised through a many body interatomic interaction model[30]. The $y$ and $z$ dimensions are $39.5a<111>/2$ and $20a<\bar{1}01>$ respectively with $a=0.31472$ nm as the lattice constant, whereas the $x$ dimension along $<1\bar{2}1>$ is varied to alter the link length. An edge dislocation is introduced with its line along the $x$ axis and the Burgers vector $a<111>/2$ along the $y$ direction. The void shown in Fig. 1a is created by removing all the atoms within a distance of 0.5 nm from the centre of the void. Periodic boundary condition is imposed along the directions of the dislocation line and the Burgers vector, while the top and bottom surfaces are kept free for the application of surface tractions. The whole system is equilibrated at 100 K and then the atoms at free boundaries are subjected to precisely calculated forces, so that a shear stress of 50 MPa is created and maintained throughout the simulation[31]. The atomic coordinates are calculated with time step of 0.5 fs using the velocity verlet integrator[31] in conjunction with the the Nosé-Hoover thermostat[32,33]. The dislocation core atoms are identified by using the suitable filtering windows based upon the centrosymmetric deviation parameter[31] and the dislocation position is defined as the centre of mass of the core atoms, which is recorded at 0.25 ps time interval. We have repeated these simulations at higher temperatures and observed enhanced damping as expected.

**Numerical Simulation of oscillating string.** The discretisation of a string is done by approximating it as a series of closely spaced beads of some specific mass, representing the linear density of the string. Each bead is connected to the neighbouring beads through massless strings under a given tension. This approximation is reasonable as it can be proved[34] that the resulting dynamics converges to that of a continuous string in the limit where the number of beads tends to infinity. Here we assume the beads of unit mass to be separated by unit distance (arbitrary units). In case of perfect boundaries, the end point coordinates remain invariant, whereas for flexible boundaries, hookean restoring forces are added. Equation (2) is numerically solved using the Velocity Verlet algorithm under the required boundary (periodic for flexible ends) and initial conditions. The viscous drag coefficient and the line tension are 0.2 and 1000 for the results presented in Fig. 4.

**Acknowledgement**   One of the authors (A. D.) thanks the Council of Scientific and Industrial Research, India, for providing financial support in the form of senior research fellowship.

**Figure Captions**

**Figure 1 Storyboard of dislocation pinning. a-f,** Snapshots of dislocation void interaction and subsequent oscillation at different time steps at 100 K. The dislocation link length is 6.9 nm. **g,** Oscillatory motion of the pinned segment is visible through the position of the centre of mass of the dislocation core atoms with respect to time. The centre of mass positions corresponding to the MD snapshots are indicated in the inset.

**Figure 2 Link length dependence of oscillation frequency. a,** Frequency spectrum of oscillation data for two different link lengths with $2^9$ data points used for each FFT analysis. The inset shows the raw data after removing the initial part up to the first oscillation peak and bringing the data to the zero mean value. The longer link length produces low frequency oscillation. **b,** $f_0^2$ as obtained from the FFTs are plotted as a function of $1/l^2$.

**Figure 3 Flexible boundaries. a,** Schematic of flexible boundary condition modelled as the ends vibrating segments attached with springs. **b,** Graphical solution of equation (8). For a given spring constant k, the locus of points of intersection of the plane $\Gamma=k$ with the plotted surface represents the allowed relations between $p$ and $l$.

**Figure 4** Numerical implementation of the flexible boundary model. **a,** A typical frequency spectrum obtained after introducing springs with $k$=10 (a.u.) in the vibrating string model with damped oscillations shown in the inset. **b,** $f_0^2$ vs. $l^{-2}$ plot with three different values of $k$. The inset shows the results for rigid boundary condition. A resemblance in the trends between Figs. 2b and 4b is noticeable. **c,** $f_0^2$ vs. $l^{-1}$ plot with different values of $k$ has been presented with linear fits. **d,** $f_0^2$ vs. $l^{-1}$ plot of the MD data with linear fit to extract the k value from slope.

**Supplimentary note 1**: Vibrating string with rigid boundary.

The equation of motion is

$$M\frac{\partial^2 y}{\partial t^2} + B\frac{\partial y}{\partial t} - T\frac{\partial^2 y}{\partial x^2} = 0, \text{ which can be written as}$$

$$\frac{\partial^2 y}{\partial t^2} + 2\eta\frac{\partial y}{\partial t} - c^2\frac{\partial^2 y}{\partial x^2} = 0, \qquad (1.1)$$

where $\eta = \dfrac{B}{2M}$ and $c = \sqrt{T/M}$

Using the variable separation method, we obtain the trial oscillatory solution

$$y(x,t) = e^{-\eta t}(C_1 \cos px + C_2 \sin px)(C_3 \cos \omega t + C_4 \sin \omega t), \qquad (1.2)$$

where $p$ is the wave vector and $\omega^2 = p^2 c^2 - \eta^2$. Rigid boundary conditions are

$$y(0,t) = 0 \text{ and } y(l,t) = 0. \qquad (1.3)$$

Using condition $(1.3)_1$ we obtain

$$y(x,t) = C_2 e^{-\eta t} \sin px (C_3 \cos \omega t + C_4 \sin \omega t), \text{ while condition } (1.3)_2 \text{ gives}$$

$$pl = n\pi, \quad n = 1,2,3,...$$

Finally we have

$$y(x,t) = \sum_{n=1}^{\infty} C_{2n} e^{-\eta t} \sin\frac{n\pi x}{l}(C_{3n} \cos\sqrt{\frac{n^2\pi^2 c^2}{l^2} - \eta^2}\, t + C_{4n} \sin\sqrt{\frac{n^2\pi^2 c^2}{l^2} - \eta^2}\, t). \qquad (1.4)$$

Arbitrary initial conditions are

$$y(x,0) = f(x) \text{ and } \left.\frac{\partial y}{\partial t}\right|_{t=0} = g(x). \qquad (1.5)$$

From (1.4) and $(1.5)_1$ we get

$$f(x) = \sum_{n=1}^{\infty} A_n \sin\frac{n\pi x}{l}, \quad A_n = C_2 C_3.$$

We compare this with the half range sine series to obtain

$$A_n = \frac{2}{l}\int_0^l f(x)\sin\frac{n\pi x}{l}dx. \quad (1.6)$$

Using (1.4), (1.5)$_2$ and (1.6) we observe

$$g(x) = \sum_{n=1}^{\infty}\left[C_{2n}C_{4n}\sqrt{\frac{n^2\pi^2c^2}{l^2}-\eta^2} - \eta B_n\right]\sin\frac{n\pi x}{l}.$$

Again from the half range sine series

$$C_{2n}C_{4n}\sqrt{\frac{n^2\pi^2c^2}{l^2}-\eta^2} - \eta A_n = \frac{2}{l}\int_0^l f(x)\sin\frac{n\pi x}{l}dx.$$

Assuming $C_{2n}C_{4n}=B_n$, we can write

$$B_n = \frac{2}{\sqrt{n^2\pi^2c^2-\eta^2 l^2}}\int_0^l[\eta f(x)+g(x)]\sin\frac{n\pi x}{l}dx. \quad (1.7)$$

Collecting equations (1.4), (1.6) and (1.7), we have

$$y(x,t) = \sum_{n=1}^{\infty}e^{-\eta t}\sin\frac{n\pi x}{l}\left(A_n\cos\sqrt{\frac{n^2\pi^2c^2}{l^2}-\eta^2}\,t + B_n\sin\sqrt{\frac{n^2\pi^2c^2}{l^2}-\eta^2}\,t\right). \quad (1.8)$$

After having formulated the generalized solution, we now consider our case study. The initial conditions compatible with the MD simulations are

$$f(x) = 0 \text{ and } g(x) = v_0[u(x) - u(x-l)].$$

Equations (1.7) and (1.8) yield

$$A_n = 0, \quad B_n = \frac{4v_0 l}{(2m+1)\pi\sqrt{(2m+1)^2\pi^2c^2-\eta^2 l^2}}, \text{ where } m = \frac{n-1}{2} = 0,1,2,... \quad (1.9)$$

From equations (1.8) and (1.9) we find

$$y(x,t) = \sum_{m=0}^{\infty}\frac{4v_0 l}{(2m+1)\pi\sqrt{(2m+1)^2\pi^2c^2-\eta^2 l^2}}e^{-\eta t}\sin\sqrt{\frac{(2m+1)^2\pi^2c^2}{l^2}-\eta^2}\,t\sin\frac{(2m+1)\pi x}{l}. \quad (1.10)$$

The centre of mass is evaluated as

$$y_{cm}(t) = \frac{1}{l}\int_0^l y(x,t)dx = \sum_{m=0}^{\infty} \frac{8v_0 l}{(2m+1)\pi\sqrt{(2m+1)^2\pi^2 c^2 - \eta^2 l^2}} e^{-\eta t} \sin\sqrt{\frac{(2m+1)^2\pi^2 c^2}{l^2} - \eta^2}\, t. \quad (1.10)$$

The most prominent is the first harmonic with $m=0$.

$$y_{cm}(t) = \frac{8v_0 l}{\pi\sqrt{\pi^2 c^2 - \eta^2 l^2}} e^{-\eta t} \sin \omega t, \quad \text{where} \quad \omega^2 = \omega_0^2 - \eta^2 \quad \text{with the natural frequency}$$

$\omega_0 = \pi c/l.$

**Supplimentary note 2**: Vibrating string with flexible periodic boundary.

We begin with the trial oscillatory solution.

$$y(x,t) = e^{-mt}(C_1 \cos px + C_2 \sin px)(C_3 \cos \omega t + C_4 \sin \omega t). \qquad (2.1)$$

The boundary conditions are

$$y(0,t) = y(l,t) \text{ [periodic boundary condition] and } T\left[\left.\frac{\partial y}{\partial x}\right|_{x=0} - \left.\frac{\partial y}{\partial x}\right|_{x=l}\right] = ky(0,t), \qquad (2.2)$$

where k is the spring constant. From equations (2.1) and (2.2)$_1$ we find

$$\frac{C_2}{C_1} = \frac{1-\cos pl}{\sin pl}, \quad pl \neq n\pi, \quad n = 1,2,3,... \qquad (2.3)$$

$pl = n\pi$ does not yield a meaningful result with equation (2.2) and we proceed with equation (2.3).

Using equations (2.1) and (2.2)$_2$ we have

$$Tp\left[\frac{C_2}{C_1}(1-\cos pl) + \sin pl\right] = kC_1. \qquad (2.4)$$

Substituting equation (2.3) in (2.4) yields

$$2pT \tan\frac{pl}{2} = k \qquad (2.5)$$

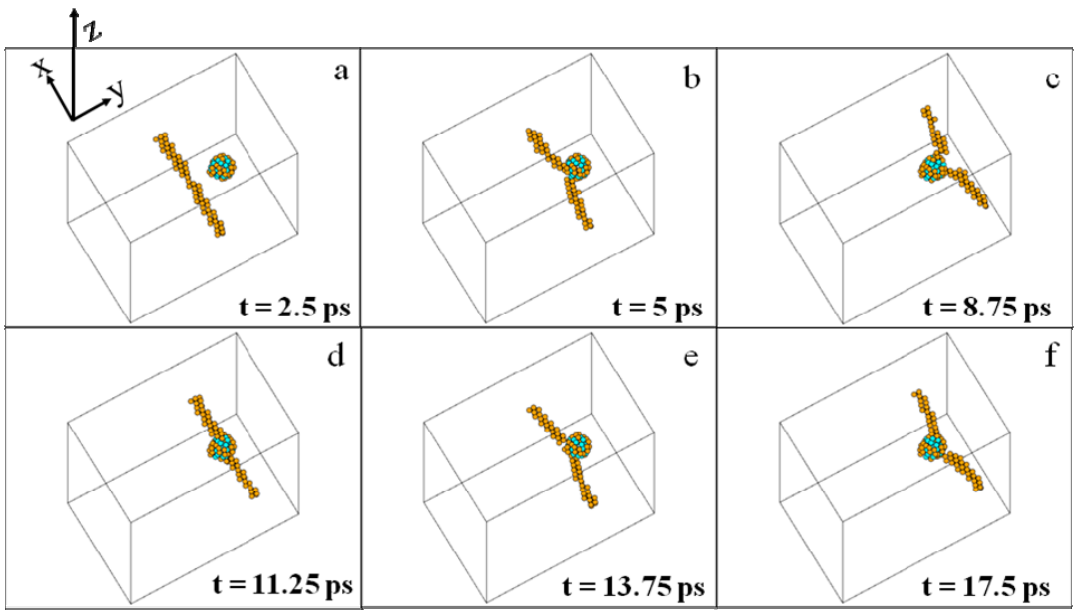

Figure 1

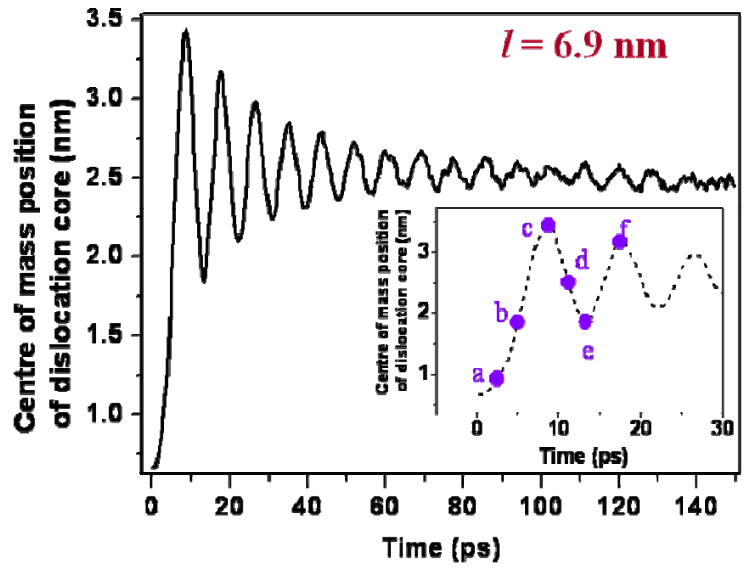

Figure 1g

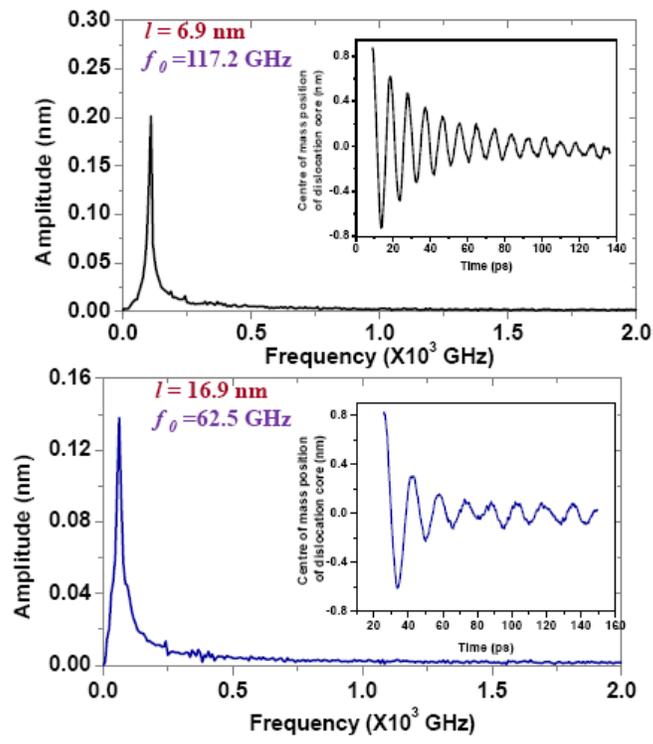

Figure 2a

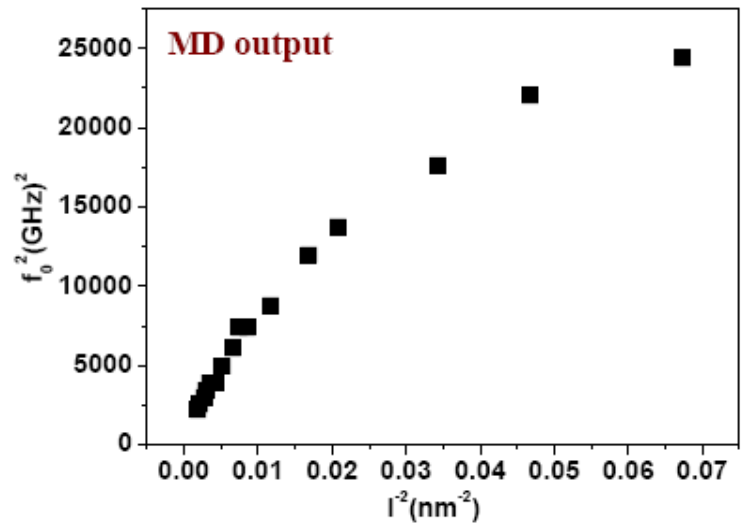

Figure 2b

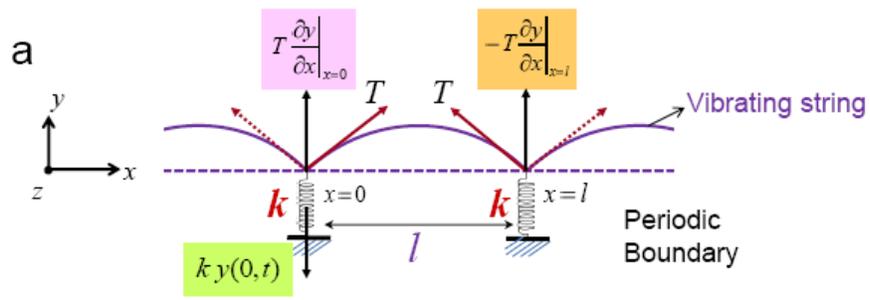

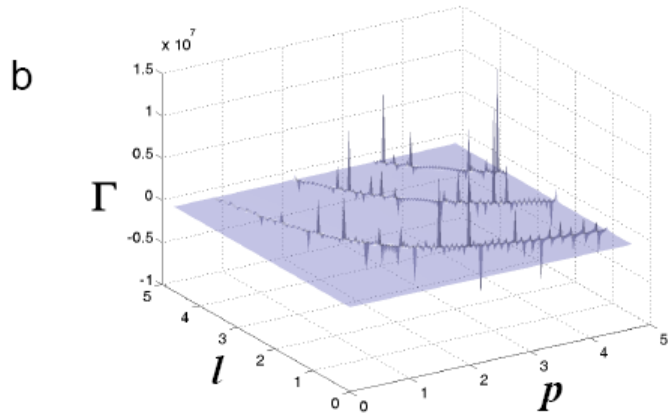

Figure 3

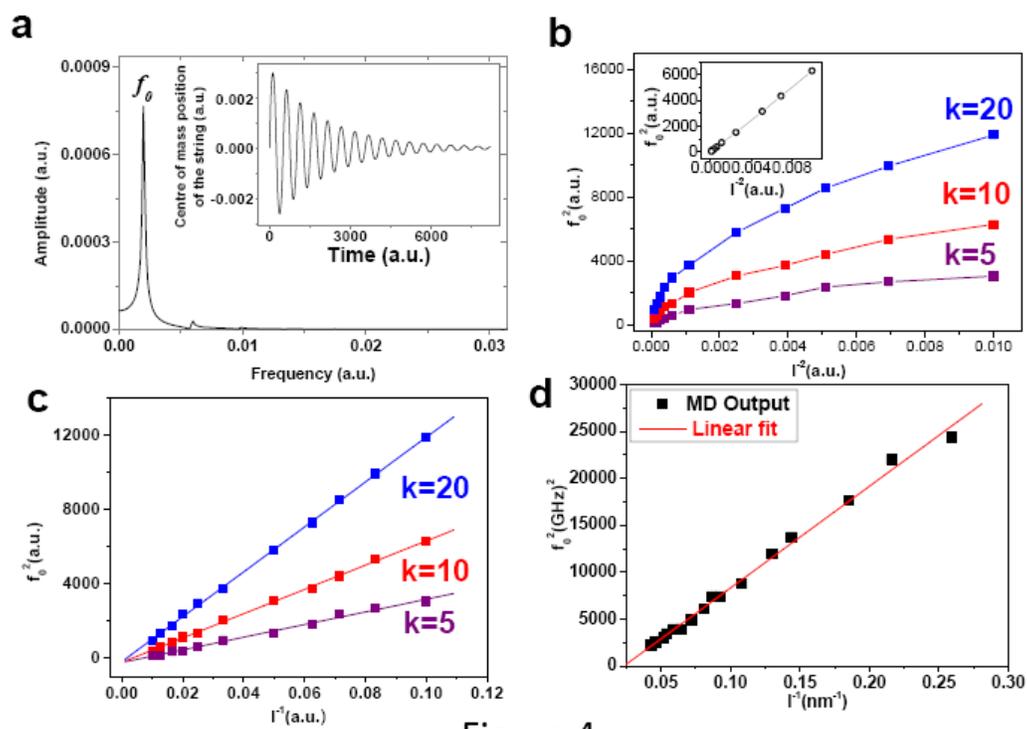

Figure 4